# Structural stability of CuAl$_2$O$_4$ under pressure

P A Agzamova[1,2], A A Belik[3] and S V Streltsov[1,2]

[1] M.N. Miheev Institute of Metal Physics of Ural Branch of Russian Academy of Sciences, 620108 Ekaterinburg, Russia
[2] Ural Federal University, Mira St. 19, 620002 Ekaterinburg, Russia
[3] International Center for Materials Nanoarchitectonics (WPI-MANA), National Institute for Materials Science (NIMS), Namiki 1-1, Tsukuba, Ibaraki, 305−0044, Japan

E-mail: polly@imp.uran.ru

## Abstract

Structural properties of CuAl$_2$O$_4$, which was recently argued to show unusual suppression of the Jahn-Teller distortions by the spin-orbit coupling, are investigated under pressures up to 6 GPa. Analysis of X-ray powder diffraction experiments shows that CuAl$_2$O$_4$ gets unstable and decomposes onto CuO and Al$_2$O$_3$ at pressures ~6 GPa and temperature ~1000 K. This finding is complemented by the DFT+U+SOC calculations, which demonstrate that this instability is partially driven by a (relatively) large compressibility of strongly Jahn-Teller distorted CuO.

## 1. Introduction

The $AB_2O_4$ ($A$ – divalent cation such as Cu, Co; $B$ – trivalent cation, e.g., Al, Fe) family of spinel oxides is known for their unique physical properties, which have been attracting considerable attention over the years. There are multiferroics among them, e.g., CdV$_2$O$_4$ [1], CoCr$_2$O$_4$ [2], and FeCr$_2$O$_4$ [3, 4], materials with charge [5] and orbital ordering [6], some of these spinels show rather unusual distortions of the crystal structure, which can be accompanied by strong modifications of their magnetic properties and even openning the spin gap [7– 9].

In fact two different sites are available for metals in spinels: octahedral and tetrahedral. In "normal" spinels – $A(B)_2O_4$ – $B$ ions sit at octahedral positions, while in "inverse" ones – $B(AB)O_4$ – they are distributed equaly between octahedral and tetrahedral voids (and of course an intermediate situation with random destribution is also possible). The transition metals (TM) can occupy both sites. While a lot of affords are now concentrated on studying the spinels with TMs having octahedral coordination, an opposite situation is equally interesting. In the present paper we study structural properties of CuAl$_2$O$_4$, where Cu$^{2+}$ ions preferably occupy tetrahedral sites. They have $3d^9$ electronic configuration and in tetrahedral surrounding Cu $3d$ orbitals are split onto three high-lying $t_{2g}$-levels separated from two low-lying $e_g$-levels. Therefore, there is one hole in the $t_{2g}$ manifold in CuAl$_2$O$_4$, which is susceptible to further Jahn-Teller distortions to lower the total energy. However, previous X-ray powder diffraction studies did not detect any indications of the Jahn-Teller effect in CuAl$_2$O$_4$ [10]. It was later shown theoretically that the absence of the Jahn-Teller distortions can be related with formation of the $j_{eff}=1/2$ state [11, 12]. Thus, it was suggested that Cu$^{2+}$ ions behave in this material similar to Ir$^{4+}$ ions also having $t_{2g}^5$ electronic configuration. Strong spin-orbit splitting lifts degeneracy and puts a single hole onto the $j_{eff}=1/2$ spin-orbital retaining only Kramers degeneracy, which can not be removed by the Jahn-Teller distortions, see e.g. [13]. Moreover, it was shown that suppression of the Jahn-Teller effect is not gradual, but there is a critical value of strength of the spin-orbit coupling above which distortions vanish [14, 15].

Possible stabilization of the $j_{eff}=1/2$ state makes CuAl$_2$O$_4$ a unique system, which is similar to famous iridates [16, 17], but based on the $3d$ transition metal Cu. First theoretical calculations indeed demostrated that one might expect strong anisotropy of the exchange coupling and stressed importance of the frustration effects, since magnetic Cu$^{2+}$ ions form strongly frustrated diamond lattice [11].

However, all these results are based on, first, the assumption that there is an ideal ordering of $Cu^{2+}$ ions and they occupy only tetrahedral sites (i.e. this is normal spinel), while there are experimental evidences of antisite defects [10, 18]. Second, there is of course always competition between the Jahn-Teller effect and spin-orbit coupling and, e.g., theoretical calculations show that applying pressure one may shift the balance between these two factors and $CuAl_2O_4$ is expected to show Jahn-Teller distortions already at pressure [12]. Third, because of antisite disorder observed in previous studies it is highly desirable to find alternative synthesis methods or ways, which would modify the distribution of $Cu^{2+}$ cations between the tetrahedral and octahedral sites of the spinel structure. The high-pressure high-temperature annealing is one of such methods.

In this work we focused on experimental and theoretical investigations of structural properties of $CuAl_2O_4$ under high pressure. Ex-situ X-ray powder diffraction studies showed that at temperatures above about 1070 K and pressure of 6 GPa $CuAl_2O_4$ decomposes through a complex way to a final mixture of initial oxides $CuO$ and $Al_2O_3$. The DFT+U+SOC calculations confirmed that $CuAl_2O_4$ is unstable at high pressure.

## 2. Experimental details and results

$CuAl_2O_4$ was prepared from a stoichiometric mixture of $Al_2O_3$ (99.9%) and $CuO$ (99.9%). The mixture was pressed into a pellet and annealed on Pt foil at 1193 K for 84 h and at 1293 K for 38 h in air with several intermediate grindings. This phase will be called ambient-pressure (AP) $CuAl_2O_4$, and it had orange-brown color. As-prepared single-phase $CuAl_2O_4$ was then annealed in Au capsules at 6 GPa and different temperatures for 2 h using a belt-type high-pressure apparatus (where the annealing temperature was reached in 10–15 min). After the high-pressure high-temperature treatments, the samples were quenched to room temperature (RT) by turning off current, and the pressure was slowly released. Such samples will be called high-pressure (HP) samples. We emphasize that the melting point of Au is above about 1600 K at 6 GPa [19]; therefore, the high-pressure annealing could be performed safely up to 1523 K in gold capsules.

X-ray powder diffraction (XRPD) data were collected at RT on a RIGAKU MiniFlex600 diffractometer using CuKα radiation ($2\theta$ range of 8–140°, a step width of 0.02°, and a scan speed of 1 deg/min). XRPD data were analysed by the Rietveld method using *RIETAN-2000* [20].

$CuAl_2O_4$ prepared at ambient pressure was single-phase and had sharp reflections on XRPD patterns. The structural analysis gave the following cation distribution $[Cu_{0.676}Al_{0.324}]_{8a}[Al_{1.676}Cu_{0.324}]_{16d}O_4$. The similar distribution was found in the literature for samples prepared at ambient pressure (for example, $[Cu_{0.68}Al_{0.32}]_{8a}[Al_{1.68}Cu_{0.32}]_{16d}O_4$ in [10] or $x = 0.36–0.39$ for $[Cu_{1-x}Al_x]_{8a}[Al_{2-x}Cu_x]_{16d}O_4$ in [18]).

XRPD patterns of the samples annealed at 6 GPa and different temperatures are given on Figure 1. The spinel structure remains after annealing at 6 GPa and 880 K (powdered sample had orange-brown color) or 1048 K (powdered sample had brown color).

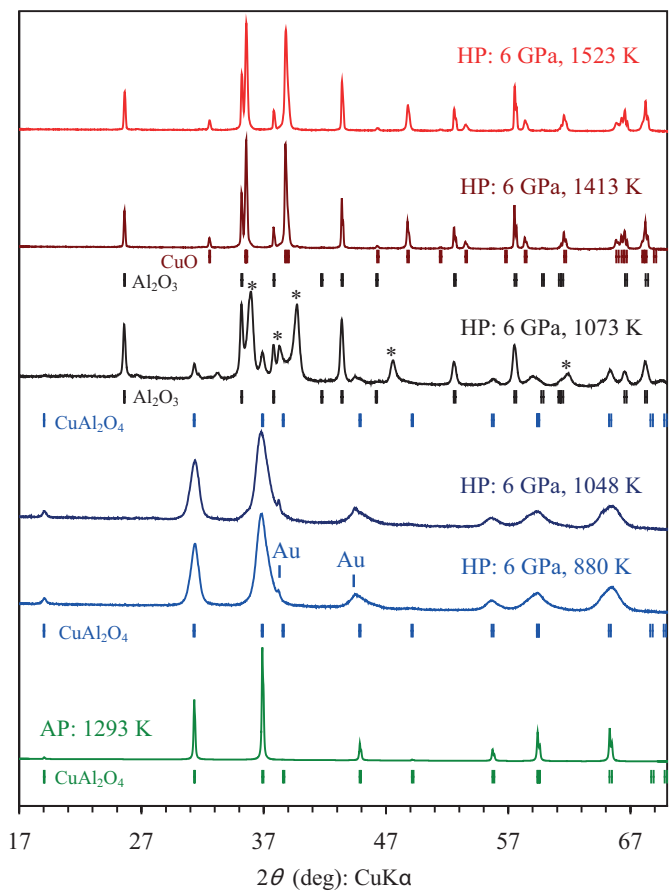

**Figure 1.** Experimental X-ray powder diffraction patterns of $CuAl_2O_4$ prepared at ambient pressure (bottom) and at 6 GPa and different temperatures (annealing conditions are given on the figure). Possible Bragg reflections for the given phases are shown by tick marks. Stars mark the strongest unidentified reflections.

However, the reflections of the spinel phase were significantly broadened, and additional reflections from Au (the capsule material) appeared. Particle size is unlikely to significantly decrease at such annealing conditions. Therefore, the broadening of the reflections could originate from the accumulation of huge stress. The appearance of Au could be caused by a small and gradual loss of oxygen or copper by $CuAl_2O_4$ near contacts with Au, where oxygen or metallic copper attacks Au and demages the capsule material.

An increase of the annealing temperature to 1073 K (at 6 GPa) resulted in a partial decomposition: an XRPD pattern showed the presence of a spinel phase, $Al_2O_3$, and reflections from unknown phase(s) (see figure 2). This powdered sample had black color, and the pellet looked like a solidified melt. Further increase of the annealing temperature to 1413 K or 1523 K resulted in the complete decomposition of CuAl$_2$O$_4$ to a mixture of initial oxides (Al$_2$O$_3$ and CuO) with sharp reflections and these samples had black color. Note that no reflections from Au were observed in these cases probably because the sample was heated relatively fast above the decomposition temperature, and CuO (and Al$_2$O$_3$) does not attack Au at the annealing conditions.

It is known that at ambient pressure, CuAl$_2$O$_4$ decomposes to a mixture of Al$_2$O$_3$ and CuAlO$_2$ above about 1470 K with the loss of oxygen [21]. The application of the high pressure prevents the loss of oxygen, as can be seen from the presence of CuO instead of Cu$_2$O, but decreases the decomposition temperature of CuAl$_2$O$_4$ from about 1470 K at ambient pressure to about 1070 K at 6 GPa.

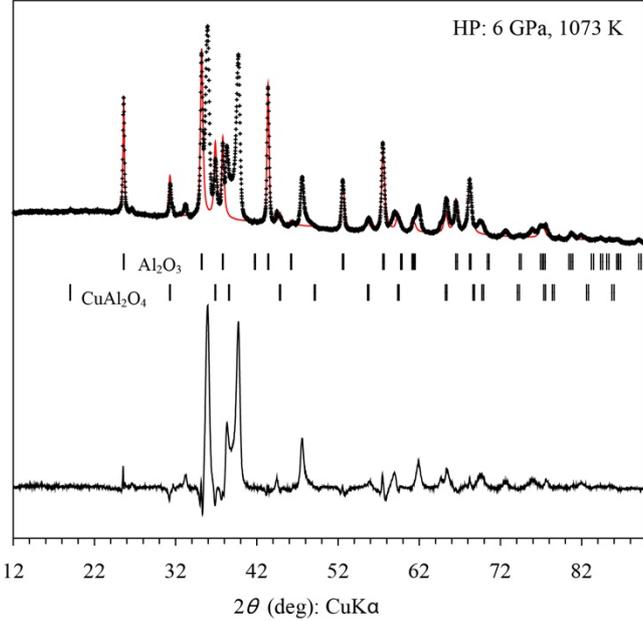

**Figure 2.** Experimental (crosses), calculated (red line), and difference X-ray powder diffraction patterns for CuAl$_2$O$_4$ annealed at 6 GPa and 1073 K. Possible Bragg reflections for Al$_2$O$_3$ and CuAl$_2$O$_4$ phases are shown by tick marks. The difference pattern emphasizes reflections from new phase(s).

## 3. Computational methods

First-principles density-functional theory (DFT) calculations were conducted to investigate the stability of CuAl$_2$O$_4$ under pressure.

The calculations of CuAl$_2$O$_4$, CuO, and Al$_2$O$_3$ compounds were carried out using the Vienna Ab initio Simulation Package (VASP) [22]. We utilized the projector augmented-wave (PAW) method [23] with the Perdew-Burke-Ernzerhof (PBE) type of exchange-correlation functional [24] within the general gradient approximation (GGA+U) [25]. Initial crystal structure parameters were taken for CuAl$_2$O$_4$ from [10] (*T=40 K*), for CuO from [26] (*T=293 K*) and for Al$_2$O$_3$ from [27]. They were relaxed for all calculated volumes. On site Coulomb repulsion parameter *U* was chosen to be 9 eV, while intra-atomic Hund's exchnage 1 eV [12]. The cut-off energy was taken to be $E_{cutoff}$=560 eV and 6×6×6 Monkhorst-Pack grid of *k*-points was used during calculations. The spin-orbit coupling (SOC) was included to the calculation scheme for CuAl$_2$O$_4$ and CuO.

Strictly speaking DFT is valid only for zero temperature, while majority of high-pressure experimental work is performed at 300 K and above. Therefore, we would like to mention that there is always at least the 300 K difference between theoretical and experimental conditions. This temperature difference is sometimes responsible for different values of pressure when a phase transition is predicted from first-principles calculations and observed in experiments.

## 4. Theoretical study of structural stability

The total energy dependence on the unit cell volume for CuAl$_2$O$_4$, CuO, and Al$_2$O$_3$ was obtained by series of calculations, where it was varied between about –10% and +10%.

First of all, we see that equlibrium volumes V$_0$ obtained in our calculations are close to experimental ones, so that $(V_0 - V_0^{exp})/V_0^{exp}$ is 0.031 for CuAl$_2$O$_4$ [10], 0.008 for CuO [26] and 0.027 for Al$_2$O$_3$ [27], which are very typical estimates for DFT [28]. The energy-volume curves for CuAl$_2$O$_4$, CuO, and Al$_2$O$_3$ compounds are given in Figure 3.

Next we extracted other parameters of the equation of state (the zero pressure total energy $E_0$, the equilibrium volume $V_0$, and the bulk modulus ($B_0$) by fitting the calculated energies versus volume to the third order Birch-Murnaghan equation of states [29]. Already at this level we see that compressibility (inverse of Bulk modulus) of CuO is much larger than in both CuAl$_2$O$_4$ and Al$_2$O$_3$ (table 1).

**Table 1.** The equation of state parameters of CuAl$_2$O$_4$, CuO, and Al$_2$O$_3$ obtained by fitting the calculated GGA+U+SOC energies versus volume data to the third order Birch-Murnaghan equation of states.

|  | CuAl$_2$O$_4$ | CuO | Al$_2$O$_3$ |
| --- | --- | --- | --- |
| $E_0$, eV | – 90.8 | – 31.2 | – 224.5 |
| $V_0$, Å | 135.2 | 81.6 | 262.9 |
| $B_0$, GPa | 187.7 | 140.4 | 230.8 |
| $B_0$' | 4.2 | 4.7 | 4.04 |

The thermodynamic stability of CuAl$_2$O$_4$ was investigated by comparing the enthalpies of CuAl$_2$O$_4$ and CuO+Al$_2$O$_3$, which are shown in Figure 4. One may see that in DFT+U+SOC calculations CuAl$_2$O$_4$ becomes unstable at pressures ~8 GPa, which is close to experimental $P_c$=6 GPa.

It is instructive to study physical origin of this decomposition. For this in Table 2 we present results of fitting of enthalpy by the third order polynomials

$$H(P) = a \cdot P^2 + b \cdot P + c + d \cdot P^3 \qquad (1)$$

for $CuAl_2O_4$, CuO, and $Al_2O_3$ compounds (here $P$ stands for pressure).

**Table 2.** Parameters of fitting GGA+U+SOC enthalpies by the third order polynomials $a \cdot P^2 + b \cdot P + c + d \cdot P^3$ for $CuAl_2O_4$, CuO, and $Al_2O_3$. Here $P$ stands for pressure.

| | $a, \frac{eV}{(GPa)^2}$ | $b, \frac{eV}{GPa}$ | $c, eV$ | $d, \frac{eV}{(GPa)^3}$ |
|---|---|---|---|---|
| $CuAl_2O_4$ | –0.001 | 0.422 | –45.378 | $6.723 \cdot 10^{-6}$ |
| CuO | –0.0005 | 0.127 | –7.808 | $4.547 \cdot 10^{-6}$ |
| $Al_2O_3$ | –0.0006 | 0.274 | –37.409 | $2.773 \cdot 10^{-6}$ |

By analyzing parameters of fitting, one can see that quadratic and cubic terms, characterized by coefficients $a$ and $d$ respectively, are much smaller than other terms (parametrized by $b$ and $c$) and therefore these coefficients only weakly affect structural stability of the materials under consideration in a given pressure range.

One can see that the constant term ($c$) is much smaller for $CuAl_2O_4$ and this reflects structural stability of this material at normal conditions.

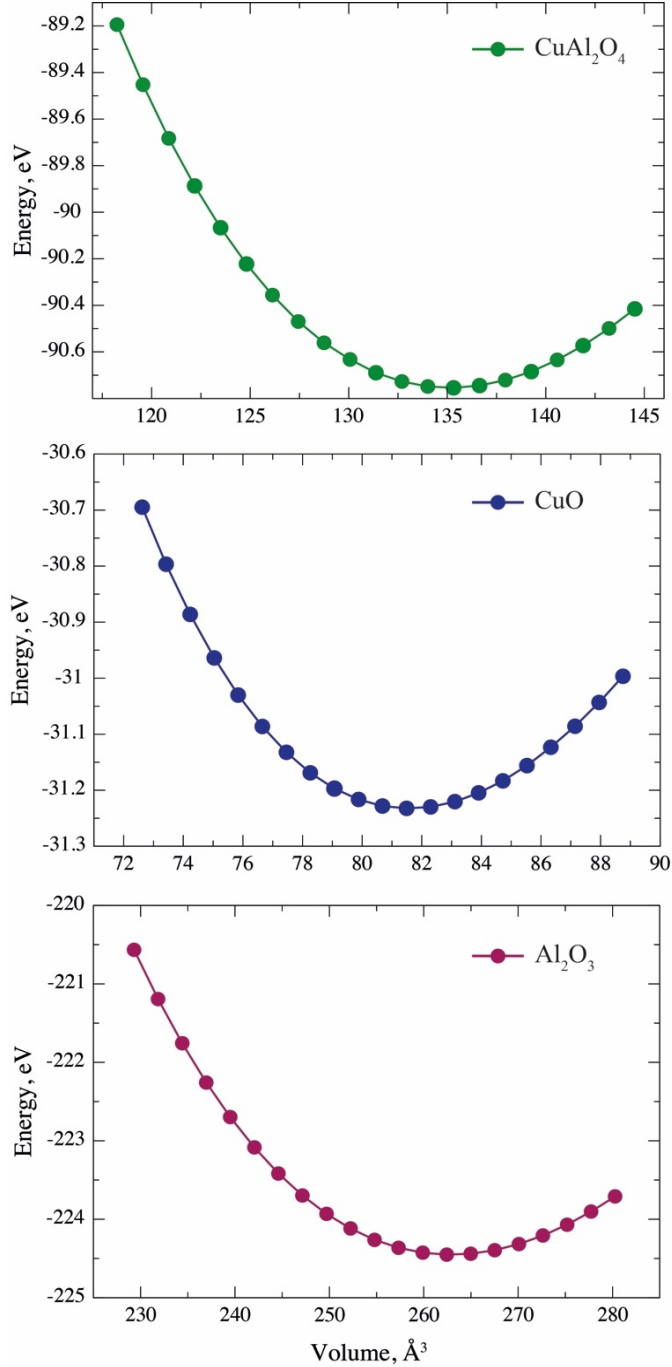

**Figure 3.** Energy-volume curves for $CuAl_2O_4$, CuO, and $Al_2O_3$ obtained by fitting the third order Birch-Murnaghan equation of states to energy-volume data.

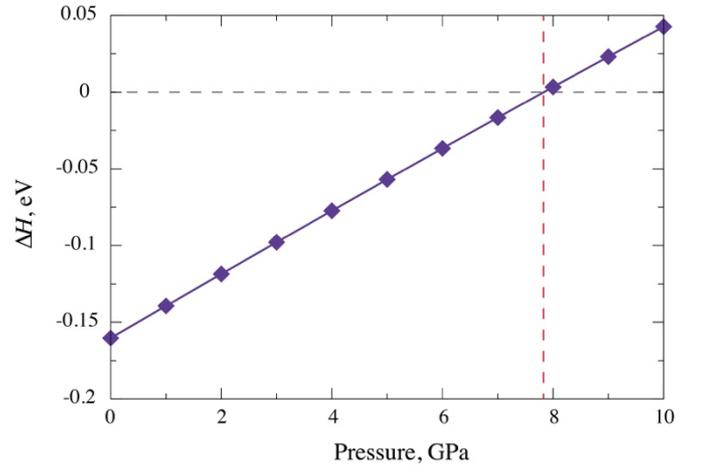

**Figure 4.** The calculated relative enthalpy ΔH of $CuAl_2O_4$ with respect to (CuO+$Al_2O_3$) under various pressure conditions. One can see that $CuAl_2O_4$ is stable below ~8 GPa.

The linear term, characterized by positive coefficient $b$, destabilizes all compounds, since the enthalpy grows in all of them due to this linear term. However, it increases very differently for $CuAl_2O_4$, CuO, and $Al_2O_3$. If $b$ for mixture of CuO and $Al_2O_3$ would be larger than for $CuAl_2O_4$ then the later would be stable at any pressure. Our calculations show that this is not the case: $b_{CuAl2O4} = 0.422$ eV/GPa, while $b_{CuO+Al2O3}$ =0.403 eV/GPa and this is one of the reasons why $CuAl_2O_4$ decomposes under pressure. One can also see that $b$ coefficient for CuO is much smaller than for $Al_2O_3$ and therefore at pressures ~8 GPa, the largest contribution to relative (with with respect to $CuAl_2O_4$) decrease of the enthalpy of the mixture CuO + $Al_2O_3$ ($\delta H_{CuO+Al2O3}$ =–3.54 eV) is mostly due to CuO ($\delta H_{CuO}$ =–2.36 eV). Thus, small $b$ for CuO coefficient is one of the most important factors, which results in decomposition of CuAl$_2$O$_4$.

Interestingly, the compressibility of CuO has been studied experimentally in Ref. [26], where it was shown that it is very different for three different Cu-O bonds.

The Cu$^{2+}$ ions in this material have octahedral surrounding, but these octahedra are so elongated, that it even seems that Cu ions are in the middle of plaquetes, not octahedra, see Figure 6.

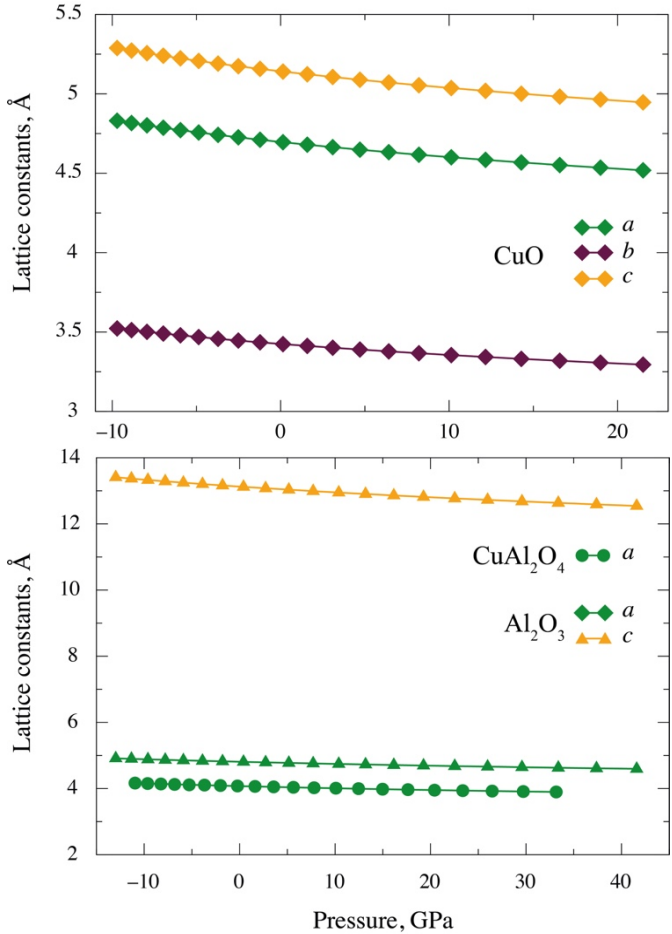

**Figure 5.** Variation of lattice parameters with pressure in CuAl$_2$O$_4$, CuO, and Al$_2$O$_3$ as obtained in the DFT+U calculations

The compressibility of long apical Cu-O bonds is much larger than others (which is rather natural) and this explains overall (relatively) large compressibility of CuO. We find at pressure 8 GPa that decrease of Cu-O bond length is only ~0.031 in CuAl$_2$O$_4$, while the longest bond in CuO changes much larger, on ~0.072 Å (two others decrease on 0.030 Å and 0.026 Å ) in the same pressure range, see figure 5.

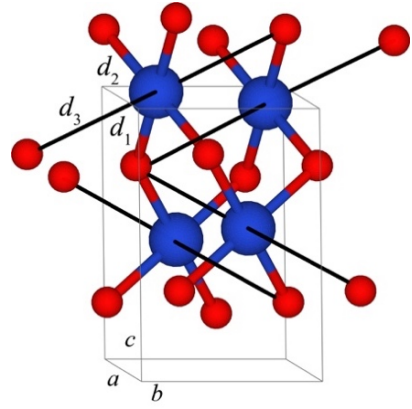

**Figure 6.** Crystal structure of CuO. The blue spheres are Cu ions, surrounded by highly distorted oxygen (the red spheres) octahedra.

The second very important factor is the electronic correlations. We found that decrease of Hubbard $U$ on 1 eV suppresses $P_c$ down to ~4 GPa, so that the optimal $U$ is ~8.5 eV. Detailed analysis shows that with decrease of $U$ nearly does not change $a, b$, and $d$ parameters in *(1)*, but decreases $c$ coeficient for CuO stronger than for CuAl$_2$O$_4$ (for $U$=8 eV $c_{CuO}$=–7.991 eV, while $c_{CuAl2O4}$=–45.779 eV). Thus, we see that correlation effects stabilize CuAl$_2$O$_4$ under pressure. This is because of weaker hybridization of Cu $3d$ states with O $2p$ states in tetrahedral geometry of CuAl$_2$O$_4$.

Of course there are also other factors, which affect decomposition of CuAl$_2$O$_4$ under pressure. Corrund ($\alpha$-Al$_2$O$_3$) is a rather stable structure: constant coeficient $a$ for Al$_2$O$_3$ is very low and this also works for stabilization of CuO and Al$_2$O$_3$ mixture. One might expect that the close packed corrundum structure is clearly more favourable than cubic packing of oxygen atoms in spinels, if we apply pressure, and a rather high temperature in our experiment also facilites the decomposition.

It has to be also mentioned that present results rely on the crystal structure with the idealized atomic order – Cu ions occupy tetrehedral sites, while Al ions are octahedral. In real CuAl$_2$O$_4$ there is a certain degree of disorder between Cu and Al and this must be taken into account in further studies.

## 6. Conclusions

To sum up the stability of CuAl$_2$O$_4$ spinel under high pressure has been investigated in the present paper experimentally and theoretically. Our ex-situ X-ray powder diffraction study showed that this material decomposes at pressure of 6 GPa above about 1070 K onto the final mixture of CuO and Al$_2$O$_3$. The DFT+U+SOC calculations generally support this finding with critical pressure of ~8 GPa for zero temperature and demonstrate that one of the important factors, which results in decomposition, is a compressibility of CuO. It is also shown that electronic correlations work in opposite direction additionally stabilizing CuO under pressure.

**Acknowledgements**

S.V.S. is grateful to I. Leonov for discussion of structural stability of normal and inverse spinels.

Theoretical calculations were supported by the Russian Science Foundation via program RSF 20-62-46047.

A.A.B. was partly supported by JSPS KAKENHI Grant Number JP20H05276, a research grant (40-37) from Nippon Sheet Glass Foundation for Materials Science and Engineering, and Innovative Science and Technology Initiative for Security (Grant Number JPJ004596) from Acquisition, Technology, and Logistics Agency (ATLA), Japan.

We also thank 02.A03.21.0006 project of Russian Ministry of Education.

**References**

[1] Giovannetti G, Stroppa A, Picozzi S, Baldomir D, Pardo V, Blanco-Canosa S, Rivadulla F, Jodlauk S, Niermann D, Rohrkamp J, Lorenz T, Streltsov S, Khomskii D I and Hemberger J 2011 *Phys. Rev. B* **83**, 060402(R)

[2] Yamasaki Y, Miyasaka S, Kaneko Y, He J-P, Arima T and Tokura Y 2006 *Phys. Rev. Lett.* **96**, 207204

[3] Singh K, Maignan A, Simon C and Martin C 2011 *Appl. Phys. Lett.* **99** 172903

[4] Eremin M V 2019 *JETP Letters* **109** 249

[5] Senn M S, Wright J P and Attfield J P 2012 *Nature* **481** 173

[6] Radaelli P 2005 *New Journal of Physics* **7** 53

[7] Schmidt M, Ratcliff W, Radaelli P G, Refson K, Harrison N M and Cheong S W 2004 *Phys. Rev. Lett.* **92** 56402

[8] Horibe Y, Shingu M, Kurushima K, Ishibashi H, Ikeda N, Kato K, Motome Y, Furukawa N, Mori S and Katsufuji T 2006 *Phys. Rev. Lett.* **96** 086406

[9] Khomskii D I and Mizokawa T 2005 *Phys. Rev. Lett.* **94** 156402

[10] Nirmala R, Jang K-H, Sim H, Cho H, Lee J, Yang N-G, Lee S, Ibberson R M, Kakurai K, Matsuda M, Cheong S-W, Gapontsev V V, Streltsov S V and Park J-G 2017 *J. Phys.: Condens. Matter* **29** 13LT01

[11] Nikolaev S A, Solovyev I V, Ignatenko A N, Irkhin V Yu and Streltsov S V 2018 *Phys. Rev. B* **98** 201106(R)

[12] Kim C H, Baidya S, Cho H, Gapontsev V V, Streltsov S V, Khomskii D I, Park J, Go A and Jin H 2019 *Phys. Rev. B* **100** 161104

[13] Streltsov S V and Khomskii D I 2017 *Phys.-Uspekhi* **60** 1121

[14] Streltsov S V and Khomskii D I 2020 *Phys. Rev. X* **10** 031043

[15] Khomskii D I and Streltsov S V arXiv:2006.05920

[16] Takagi H, Takayama T, Jackeli G, Khaliullin G and Nagler S E 2019 *Nat. Rev. Phys.* **1** 264

[17] Winter S M, Tsirlin A A, Daghofer M, van den Brink J, Singh Y, Gegenwart P and Valenti R 2017 *J. Phys.: Cond. Matt.* **29**, 493002

[18] O'Neill H, James M, Dollase W A and Redfern S A T 2005 *Eur. J. Miner.* **17** 581

[19] Decker D L and Vanfleet H B 1965 *Phys. Rev.* **138** A129

[20] Izumi F, Ikeda T 2000 *Mater. Sci. Forum* **321–324** 198

[21] Jacob K T and Alcock C B 1975 *J. Am. Ceram. Soc.* **58** 192

[22] Kresse G and Hafner J 1993 *Phys. Rev. B* **47** 558,

[23] Kresse G and Joubert D 1999 *Phys. Rev. B* **59** 1758

[24] Perdew J P, Burke K and Ernzerhof M 1996 *Phys. Rev. Lett.* **77** 3865

[25] Liechtenstein A I, Anisimov V I and Zaanen J 1995 Phys. Rev. B **52** R5467

[26] Ehrenberg H, McAllister J A, Marshall W G and Attfield J P 1999 *J. Phys.: Condens. Matter* **11** 6501

[27] Newnhan R E and de Haan Y M 1962 *Zeitschrift fur Kristallographie* **117** 235

[28] Martin R M 2004 *Electronic Structure: Basic Theory and Practical Methods* (Cambridge : Cambridge University Press)

[29] Birch F, 1947 *Phys. Rev.* **71** 11